\documentclass[lettersize,journal]{IEEEtran}                                   

\usepackage{amsmath} 
\usepackage{amssymb}  
\usepackage{mathtools}
\usepackage{multirow}
\usepackage{url}
\usepackage{float}
\usepackage{authblk}

\title{\LARGE \bf
Supertoroid fitting of objects with holes for robotic grasping and scene generation
}
\author[1]{Joan Badia Torres}
\author[1]{Eric Carmona}
\author[2]{Abhijit Makhal}
\author[3]{Omid Heidari}
\author[4]{Alba Perez Gracia}
\affil[1]{Department of Mechanical Engineering, Polythechnic University of Catalonia (UPC), Barcelona, Spain}
\affil[2]{Locus Robotics, Pittsburgh, PA}
\affil[3]{Nauticus Robotics, TX}
\affil[4]{Department of Mechanical Engineering,Polythechnic University of Catalonia (UPC), Barcelona, Spain, and \\
Institut de Robòtica i Informàtica Industrial (CSIC-UPC), Barcelona, Spain. {\tt\small alba.perez.gracia@upc.edu}}

\begin{document}

\maketitle
\thispagestyle{empty}
\pagestyle{empty}

\begin{abstract}

One of the strategies to detect the pose and shape of unknown objects is their geometric modeling, consisting on fitting known geometric entities. Classical geometric modeling fits simple shapes such as spheres or cylinders, but often those don't cover the variety of shapes that can be encountered. For those situations, one solution is the use of superquadrics, which can adapt to a wider variety of shapes. 

One of the limitations of superquadrics is that they cannot model objects with holes, such as those with handles. This work aims to fit supersurfaces of degree four, in particular supertoroids, to objects with a single hole. Following the results of superquadrics, simple expressions for the major and minor radial distances are derived, which lead to the fitting of the intrinsic and extrinsic parameters of the supertoroid. The differential geometry of the surface is also studied as a function of these parameters. The result is a supergeometric modeling that can be used for symmetric objects with and without holes with a simple distance function for the fitting.  The proposed algorithm expands considerably the amount of shapes that can be targeted for geometric modeling.

\end{abstract}

\section{INTRODUCTION}

The identification of the pose and geometry of unknown objects is needed in robotic interaction with the environment, whether to create scenes or to define objects that the robot must act on. 
Robotic picking of objects is a part of most robotics operations. A variety of picking systems have been developed to target different situations, from very general to adapted to specific situations. In any case, all of them must include, implicit or explicitly, the identification of the object or area to pick, grasp synthesis and path planning. For explicit methods, current strategies for identifying objects or areas for grasping are divided in three wide areas: model-based, geometric modeling of objects, and model-free grasping surface identification \cite{tenPas:2017}, \cite{Zapata:2019}. The geometric modeling is used also for segmentation or as a first step for model-based systems, or within machine learning algorithsm \cite{Sircelj:2020}, \cite{Tomasevic:2022}. 

Different approaches have been tested for geometric modeling fitting, from implicit polynomial and algebric invariants \cite{Keren:1996} to spherical harmonics \cite{Galin:2007}, while more recently, the fitting of simple shapes such as spheres or cylinders \cite{Chuang:2004}, \cite{Castellani:2019} is preferred,. However, those are often not similar enough to the targeted objects. 

For those situations, efforts have been developed to use superquadrics \cite{Jaklic:2000}, \cite{Saito:2001}, \cite{Vincze:2007}, \cite{Dubey:2013}, which can adapt to a variety of shapes. Superquadrics are now seeing an increase in their applications to capture point cloud geometry \cite{Menendez:2024} or to create virtual environments \cite{Paschalidou:2019}. However, superquadrics have two basic limitations: one of them is that the objects need to have axial symmetry. The second one is that they can only model objects of genus zero, that is, without holes. 

In order to overcome the first limitation, some formulas have been develop that allow the streching and deforming of superquadrics. A recent application of this can be found in \cite{KimPark:2023}. To our knowledge, these have not been used yet to fit objects for grasping. The second limitation is the one that we aim to overcome. Given the fact that many objects that robots need to grasp, especially those in the household, have holes such as handles or rings, it is important to be able to easily identify those in position and shape, in order to improve the grasping operation. However this handlers rarely have a perfect round shape, so they may be hard to fit to a regular torus.

This work aims to fit a supersurface of degree four, a superquartic with genus one, to objects with holes. In order to do that, the equations for the supertoroid are presented. 

Point clouds for supertoroids are generated using a Pilu-Fisher algorithm \cite{PiluFisher:1995}. Formulas for calculating major and minor radial distances are derived and tested. Inspired by the algorithms used in superquadrics, we obtain simple expressions for the major and minor radial distances for the fitting of the intrinsic and extrinsic parameters of the supertoroid. The result is a supergeometric modeling with a single function that can be used for symmetric objects with and without holes.  

In general, the 3D sensors are noisy and obtain only a partial view of the object, from a single viewpoint. Recent literature suggests that using a  single-view point cloud to fit superquadrics on can lead to erroneous shape and pose estimation, and multiple views are used by some authors \cite{Alaniz:2023}.  The supertoroid fitting is proved to work for partial clouds, that is, for point clouds obtained with a single camera view, under certain hypotheses, avoiding the need for point cloud completion \cite{Peters:2012}, \cite{Quispe:2015}. 

The fitting of the supertoroid allows for the easy development of heuristics for grasp synthesis. Preferred grasping points for known geometric objects such as cylinders or spheres save computational time for a known shape, see for instance \cite{Haschke:2021}. In this work, optimal grasping points can be selected as a function of the parameters of the supertoroid. In addition, the distance formula developed adds geometric meaning with respect to the signed distance formula \cite{Liu:2023}.

The proposed algorithm expands the previous work by some of the authors \cite{Makhal:2018} by increasing considerably the amount of shapes that can be targeted for geometric modeling, and opens the field to a completely new type of objects.

\section{SUPERQUARTIC SURFACES}

Quartic surfaces include a big variety of shapes. For our study, we focus on a particular rational quartic, the torus, whose equation, in a particular coordinate frame, is
\begin{equation}
    (x^2 + y^2 + z^2 + R^2 - r^2)^2 - 4R^2(x^2 + y^2) = 0,
\end{equation}
with $R$ and $r$ being the major and minor radii, respectively. 

We can generalize these quartics to toroids of different cross-sections, and to more general shapes having a single hole, if we follow the same approach defined for the superquadrics. The shape is then controlled by two exponents, $\epsilon_1$ and $\epsilon_2$, yielding the so-called \emph{supertoroids}. 

\begin{figure}[htpb]
      \centering
      \includegraphics[scale=0.22]{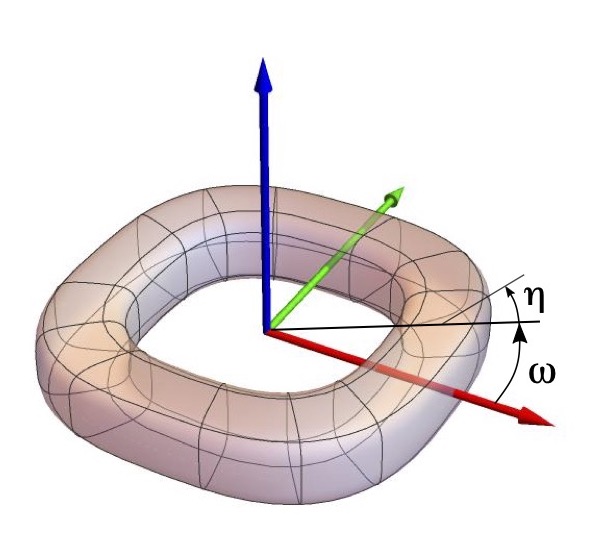}
      \caption{The angle $\omega$ on the $x-y$ plane and the angle $\eta$ for each cross-section.}
      \label{fig:stangles}
   \end{figure}

Supertoroids were already identified in the classic paper by Barr \cite{Barr:1981}, which considered both superellipsoids and supertoroids as superquadrics, given the similarity in their parametric equations. The supertoroid has both implicit and parametric expressions, as its rational quadratic counterparts. The parametric expression of the supertoroid is given by the spherical product
\begin{align}
r_t(\eta, \omega) =& \begin{Bmatrix} a_4+\cos^{\epsilon_1}\eta \\ a_3\sin^{\epsilon_1}\eta\end{Bmatrix} \otimes
\begin{Bmatrix}a_1\cos^{\epsilon_2}\omega \\ a_2\sin^{\epsilon_2}\omega \end{Bmatrix}  \nonumber \\
=& \begin{Bmatrix} a_1(a_4+\cos^{\epsilon_1} \eta) \cos^ {\epsilon_2} \omega \\ a_2(a_4+\cos^{\epsilon_1} \eta) \sin^ {\epsilon_2} \omega \\a_3\sin ^{\epsilon_1} \eta \end{Bmatrix},
\label{eq:torusParam}
\end{align}
where in order to complete the four quadrants, we need to take $\cos^{\epsilon_1} \eta = sign(\cos \eta)|\cos\eta|^{\epsilon_1}$, and similarly for the rest of trigonometric variables. Figure \ref{fig:stangles} shows the angular parameters on the supertoroid. 

\begin{figure}[htpb]
      \centering
      \includegraphics[scale=0.17]{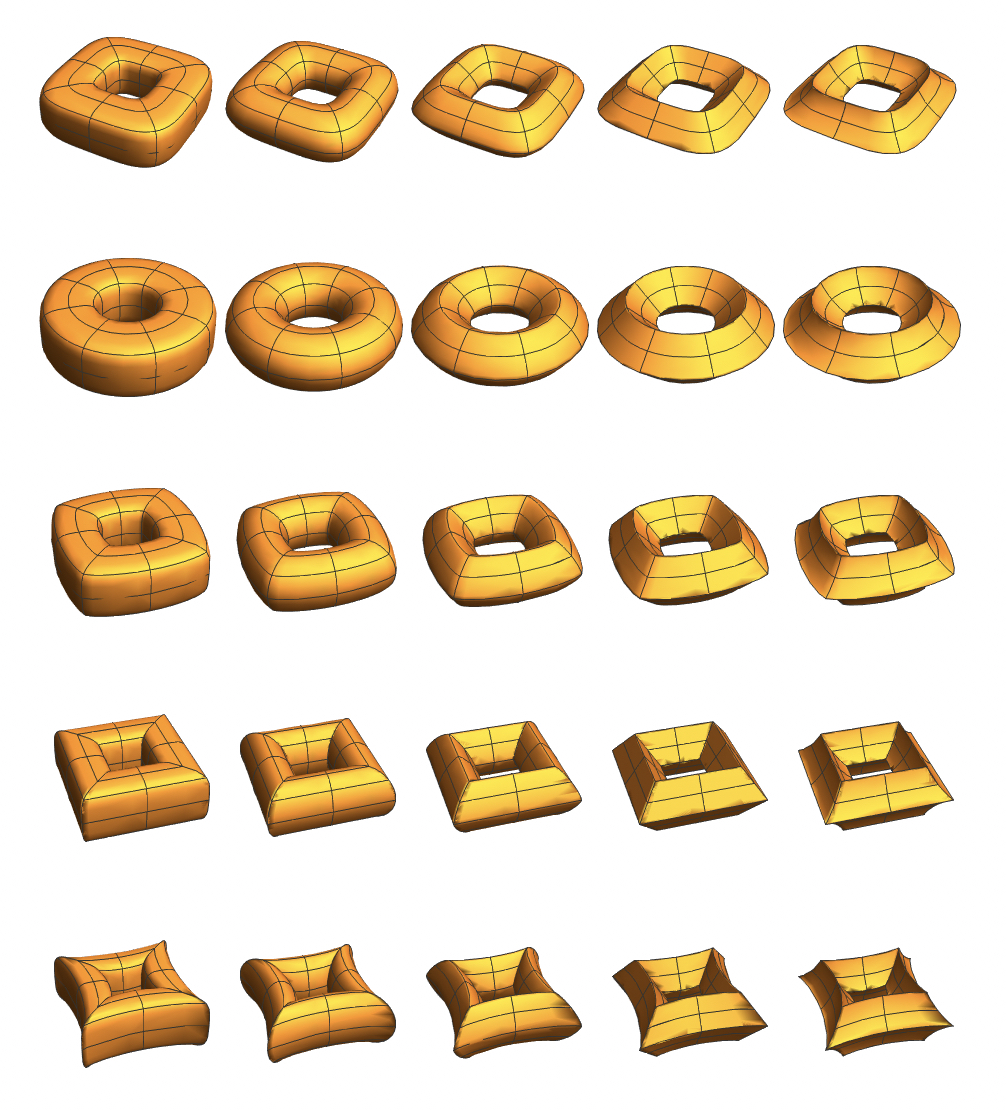}
      \caption{Different supertoroid shapes obtained with same $a_i$ values and increasing numerical values for the exponents $\epsilon_i$, from 0.5 to 2.5.}
      \label{fig:superRange}
   \end{figure}

As in the case of the superellipsoid, the real positive exponent $\epsilon_2$ controls the shape in the canonical $x-y$ plane and the real, positive exponent $\epsilon_1$ controls the shape in the perpendicular planes containing the $z$ axis. The coefficients $a_i$ control the dimensions in the $x$, $y$, $z$ directions. We collect these intrinsic parameters in a vector $v_i = (a_1, a_2, a_3, a_4, \epsilon_1, \epsilon_2)$. Fig. \ref{fig:superRange} shows a sample of shapes that can be accomplished by modifying the exponents, from smaller to larger values.

The supertoroid has the implicit equation, in its canonical coordinate system,
\begin{equation}
f_t(v_i,x,y,z): \quad \bigg|\bigg(\bigl\lvert\frac{x}{a_1} \bigl\lvert^\frac{2}{\epsilon_2} +\bigl\lvert\frac{y}{a_2} \bigl\lvert^\frac{2}{\epsilon_2} \bigg)^\frac{\epsilon_2}{2}-a_4\bigg|^{\frac{2}{\epsilon_1}} + \bigl\lvert\frac{z}{a_3}\bigl\lvert^\frac{2}{\epsilon_1} = 1.
\label{eq:toursImp}
\end{equation}
The supertoroid is bounded by planes $x=\pm(a_1+a_1a_4)$, $y=\pm(a_2+a_2a_4)$ and $z=\pm a_3$.

The implicit equation of the supertoroid defines an inside-outside function, as identified by \cite{Barr:1981}. Consider the left side of Eq.(\ref{eq:toursImp}), let us call that $F_t(v_i,x,y,z)$. Then we have the following result for points $(x,y,z)$ with respect to the superquartic:
\begin{equation}
F_t(v_i,x,y,z)  \begin{cases} 
      <1 & \textrm{inside}\ \\
     =1 & \textrm{on surface}\ \\
      >1 & \textrm{outside}\ 
   \end{cases}
\end{equation}

For our calculations, we need to define some more geometry on the supertoroid, which is detailed below. 

\begin{figure}[h!]
    \centering
    \includegraphics[scale=0.4]{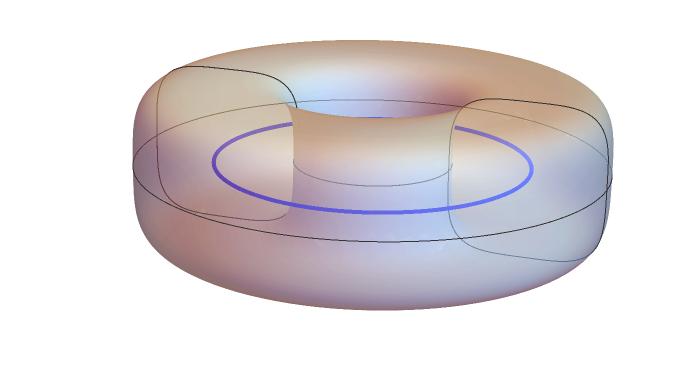}
    \caption{The mean superellipse can be seen as the mean value of the section of the supertoroid with the plane $z=0$.}
    \label{fig:STwithMSE}
\end{figure}

The center of the supertoroid for each vertical section defines a superellipse $f_m$, called the \emph{mean superellipse}, which is important for our calculations. This superellipse is located in the $x-y$ canonical plane and has implicit equation

\begin{equation}
    f_m(v_i, x, y): \quad \lvert\frac{x}{a_1 a_4} \bigl\lvert^\frac{2}{\epsilon_2} +\bigl\lvert\frac{y}{a_2 a_4} \bigl\lvert^\frac{2}{\epsilon_2} =1,
\label{eq:meanSuperellipse}
\end{equation}
assuming $a_4 > 0$. The mean superellipse is shown in Figure \ref{fig:STwithMSE}. From the parametric equation it can be seen that the mean superellipse, of parametric equation
\begin{equation}
    r_m(\omega) = \begin{Bmatrix} a_1 a_4\cos^{\epsilon_2}\omega \\ a_2 a_4\sin^{\epsilon_2}\omega\end{Bmatrix},
    \label{eq:paraMeanSE}
\end{equation}
reduces to zero when $a_4 = 0$ and both surfaces merge (for $0 < a_4 < 1$ the figure takes an apple shape). This is important when using a single equation to fit both superellipsoids and supertoroids with a general equation.

From the expression of Eq. (\ref{eq:paraMeanSE}), the radius of the mean superellipse can be derived as a function of the angle $\omega$,
\begin{equation}
    \bar R(\omega) = a_4 \sqrt{a_1^2 \cos^{2\epsilon_2}\omega + a_2^2 \sin^{2\epsilon_2}\omega},
    \label{eq:radius}
\end{equation}
in which signs to account for the different quadrants must be added when necessary.

\section{SUPERTOROID FITTING}

In order to find the supersurface that better fits a point cloud, it is necessary to define and calculate some sort of distance between each point and the surface. 

Here we derive a modification of the radial Euclidean distance studied in \cite{Whaite:1991}. In that work, the non-uniqueness of the fitting according to different metrics was also studied.  

The radial distance used in superquadrics is obtained as the distance between a point and the surface along the line joining this point to the origin, in the canonical frame of the superquadric. This distance yields an easy expression when using the implicit equation. The radial distance definition is not directly applicable to supertoroids, as the vector from the origin of the canonical frame to the point needs not intersect the surface. Hence, this definition needs to be modified for its use in supertoroids.  
  
\begin{figure}[htpb]
      \centering
      \includegraphics[scale=0.2]{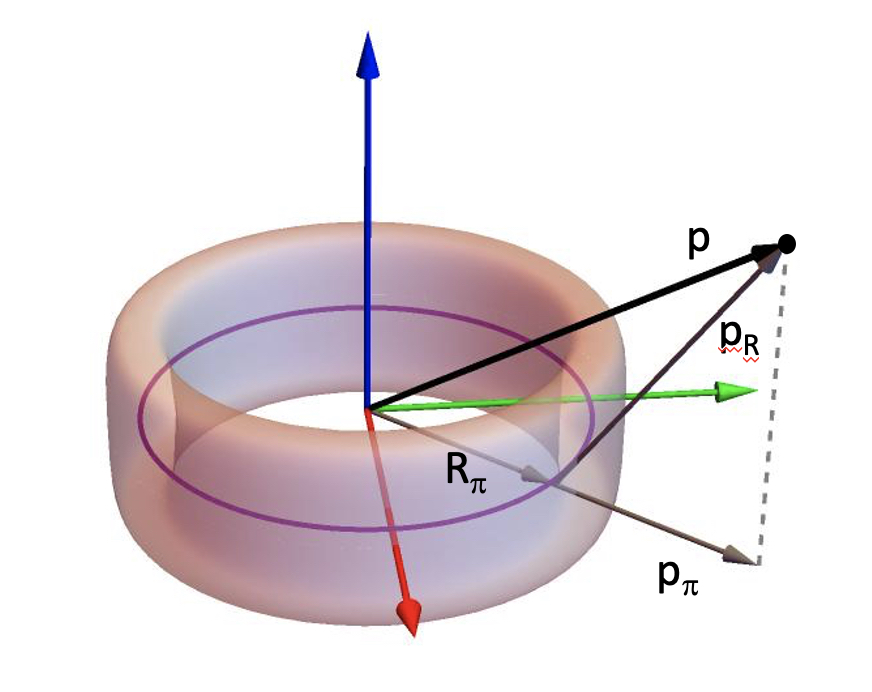}
      \caption{The vector to a point $p$ is the sum of vector $R_\pi$ (intersection of the mean superellipse with the projection of $p$ on plane $x-y$) and vector $p_R$}
      \label{fig:supertoroidVectors}
   \end{figure}

A point $p$ of the point cloud is projected in the $x-y$ plane of the supertoroid canonical frame as vector $p_\pi$. This vector will intersect the mean superellipse at a point defining the vector $R_\pi$. From that, we can create the vector $p_R$ such that $p = R_\pi + p_R$. See Fig. \ref{fig:supertoroidVectors}.

\begin{figure}[htpb]
      \centering
      \includegraphics[scale=0.17]{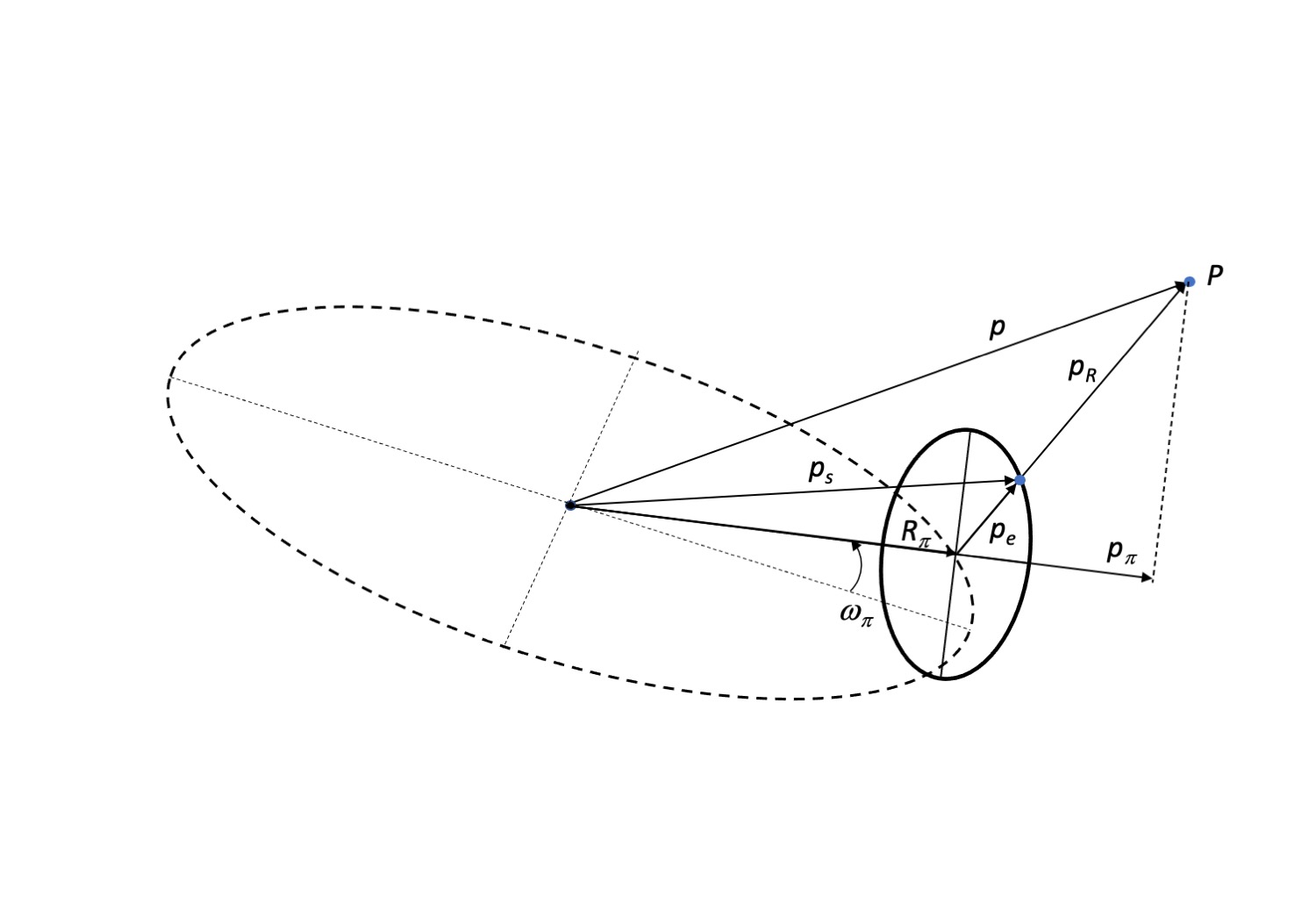}
      \caption{The vector $p_s$ of the closest point on the surface, according to the major and minor radial distance coefficients.}
      \label{fig:supertoroidVectors2}
   \end{figure}

We define two ratios, the minor ratio $\beta_1$ and the major ratio $\beta_2$. The major ratio $\beta_2$ is used to locate the intersecting point on the mean superellipse with the plane containing the vector $p$ and the $z$-axis of the superellipsoid, defining vector $R_\pi$. The minor ratio $\beta_1$ locates the tip of vector $p_e$ as a fraction of vector $p_R$, on the cross-section superellipse, see Fig. \ref{fig:supertoroidVectors2}.

The vertical projection plane is located at an angle $\omega_{\pi}$ from the canonical $x$-axis,
\begin{equation}
    \tan \omega_{\pi} = \frac{p_y}{p_x}.
\end{equation}
Considering the mean superellipse, we have the value of vector $R_\pi$ from Eq. \ref{eq:radius}. The major ratio is the fraction of the projected vector $p_\pi$ that corresponds to the length of the mean radius $R_\pi$,
\begin{equation}
    R_\pi = \beta_2 p_\pi = \beta_2 (p - (p\cdot z) z), 
\end{equation}
The coefficient $\beta_2$ can be calculated using the inside-outside function of the mean superellipse for $\beta_2 \ge 0$,
\begin{equation}
    F_m(R_\pi) = \beta_2^{2/\epsilon_2} F_m(p_\pi) = 1, .
\end{equation}

The minor ratio coefficient yields the vector $p_e$ as a fraction of the length of vector $p_R$,
\begin{equation}
    p_e = \beta_1 p_R.
\end{equation}

In order to calculate coefficient $\beta_1$ we need the expression of the superellipse $f_c$ created at each vertical cross section of the supertoroid. It is important to realize that, depending on the values of the parameters of the supertoroid, the angle of the parameterized expression at that cross section, $\omega_s$, will not correspond to the angle $\omega_\pi$ shown in Figure \ref{fig:supertoroidVectors2}. The relation between these two angles is given by
\begin{equation}
    \tan^{\frac{1}{\epsilon_2}} \omega_s = \frac{a_2}{a_1} \tan \omega_\pi.
\end{equation}

The cross-section superellipse $f_c$ can be expressed in a local canonical coordinate frame $\{x^{'},z^{'}\}$, whose orientation corresponds to a $z$-rotation by an angle $\omega_\pi$, $[R_z(\omega_\pi)]$. 

The implicit expression of the cross-section superellipse in this local frame is
\begin{equation}
    f_c(v_i, x, y, z): \quad \lvert\frac{x^{'}}{a_{\omega_s}} \bigl\lvert^\frac{2}{\epsilon_1} +\bigl\lvert\frac{z^{'}}{a_3} \bigl\lvert^\frac{2}{\epsilon_1} =1,
\label{eq:sectionSuperellipse}    
\end{equation}
where $a_{\omega_s}$ is given by
\begin{equation}
    a_{\omega_s} = \sqrt{a_1^2 \cos^{2\epsilon_2}\omega_s + a_2^2 \sin^{2\epsilon_2}\omega_s}.
\label{eq:aSubOmega}    
\end{equation}
The change of coordinates to the global frame allows us to express the local coordinates $\{x^{'},z^{'}\}$ as
\begin{align}
    x^{'} =& \frac{a_1\cos^{\epsilon_2}\omega_s(x-a_1a_4\cos^{\epsilon_2}\omega_s)}{a_{\omega_s}}+ \nonumber \\
    &\frac{a_2\sin^{\epsilon_2}\omega_s(y-a_2a_4\sin^{\epsilon_2}\omega_s)}{a_{\omega_s}} \nonumber \\
     z^{'} =& z
\end{align}

Using the inside-outside function $F_c$, we can determine the value of $\beta_1$ as a function of the point and the parameters of the superellipse. The vector $p_e$ belongs to the surface of the cross-section superellipse, so that, in local coordinates, $F_e(p_e) = F_e (\beta_1 p_R) = 1$. Using the transformation to the coordinates of the supertoroid, this yields the following equation, that allows us to calculate the minor ratio,
\begin{equation}
    \beta_1^{2/\epsilon_1} F_e(p_R) = 1.
\end{equation}

Finally, we can write the vector $p_s$ on the surface of the supertoroid as
\begin{equation}
    p_s = \beta_2 p_\pi + \beta_1 p_R = \beta_1 p + \beta_2 (1-\beta_1) p_\pi  
\end{equation}

The length of the distance between these two vectors is what we call the \emph{meridian radial distance},
\begin{equation}
    d_s = |p-p_s|.
\end{equation}
The meridian radial distance can be used for regular surfaces of revolution for which the generatrix is a closed curve. The radial vector is used to define the plane of the meridian curve, and then a line is traced between the point of the cloud and the origin of the local coordinate frame of the meridian curve, finding the distance on this line (Figure \ref{fig:supertoroidVectors2}). This distance takes a very simple expression as a function of the minor and major coefficients $\beta_1$ and $\beta_2$,
\begin{equation}
    d_s = \lvert p - p_s \lvert = \lvert (1-\beta_1)(p - \beta_2 p_\pi) \lvert.
    \label{eq:distance}
\end{equation}

The fitting procedure consists on minimizing the meridian radial distance from the point cloud to the supertoroid. This distance has been proved to give fast fitting results, as it is shown in the results section, with the additional advantage of having a geometrical meaning.

\section{SURFACE PROPERTIES FOR THE GRASPING OF SUPERTOROIDS}
Once the point cloud is fitted to a supertoroid, its twelve parameters (six extrinsic parameters for the pose of the object and six intrinsic parameters for the shape) give us all the geometric information needed for the affordance of the supertoroid, in particular for optimizing grasping strategies.

The extrinsic parameters define the local object frame. The center of the supertoroid defines the origin of the local reference frame, and the three coordinate axes correspond to the major, minor and vertical axes of the figure.

The definition of the supertoroid's geometry with these twelve parameters allows creating some heuristics for the grasping, which are dependent on the supertoroid parameters but also on the type of gripper used. 

The heuristics could be based on minimizing wrenches, especially moments, and maximizing the surface of contact between the object and the active surfaces of the gripper. In any case, finding the grasping points according to the geometry of the supertoroid implies calculating some local geometric parameters.

Even though the supertoroid is not a regular surface globally, we characterize its local differential geometry in the interval $0 <\omega, \eta < \pi/2$ range for its two parameters, avoiding the potential critical points and extending the results to the seams and stitches, and to the other quadrants given the symmetry of the surface. For background on differential geometry see \cite{Abate:2012} or \cite{DoCarmo:1976}.

\subsection{Tangent plane}

The differential geometry calculations in this and next section are done under the hypotheses of $\epsilon_1<2$ and $\epsilon_2<2$; for bigger values of the exponents, the surface presents cusps, which are treated separately. Also we assume $a_4>1$, which is necessary to have a hole in the surface. Smaller values of $a_4$ yield apple-shaped surfaces with internal tangent and normal vectors.

\begin{figure}[htpb]
      \centering
      \includegraphics[scale=0.2]{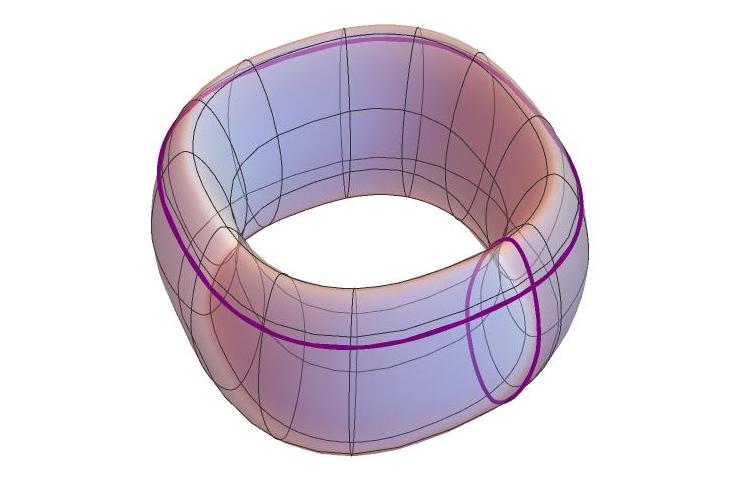}
      \caption{The coordinate curves along parameters $\omega$ and $\eta$.}
      \label{fig:coordCurves}
   \end{figure}

We use the \emph{coordinate curves} on the supertoroid defined by the two angles $\omega$ and $\eta$ shown in Figure \ref{fig:stangles}, which parameterize the surface. Those curves can be seen in Figure \ref{fig:coordCurves} and intersect but are not orthogonal in general. 

\begin{figure}[htpb]
      \centering
      \includegraphics[scale=0.3]{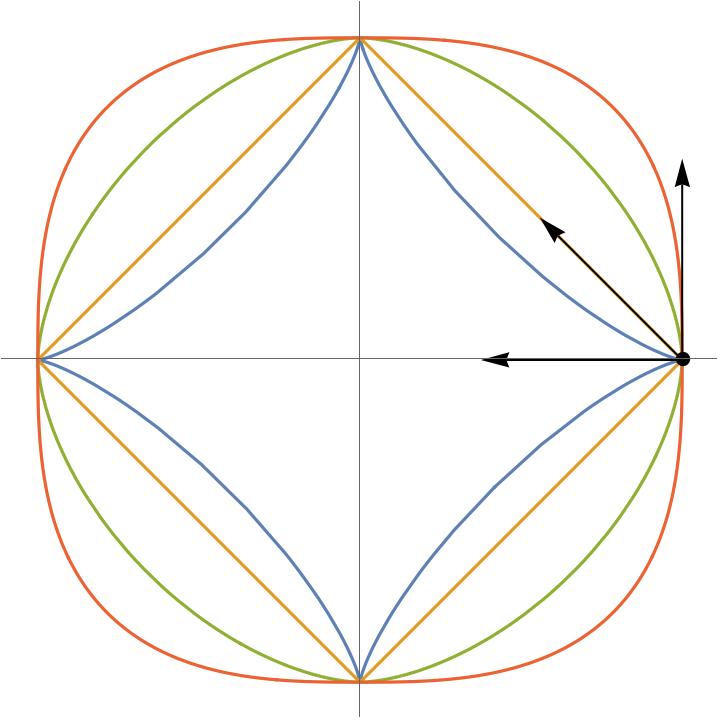}
      \caption{The tangent along the coordinate curves $\alpha_\omega$ and $\alpha_\eta$ shows similar behavior at the $\omega=0$ and $\eta=0$ points, depending on the values of $\epsilon_2$ and $\epsilon_1$. Shown are the $\epsilon_2>2$ (cusp), $\epsilon_2=2$ (singular point), and $\epsilon_2<2$}
      \label{fig:tangentLimitsOmega}
   \end{figure}

The tangents (non-unit) along those curves are
\begin{equation}
    t_{\omega} = \begin{Bmatrix} -a_1 \epsilon_2 (a_4 + 
    \cos^{\epsilon_1}{\eta})\sin{\omega} \cos^{\epsilon_2-1}{\omega} \\
    a_2 \epsilon_2 (a_4 + 
    \cos^{\epsilon_1}{\eta}) \cos{\omega} \sin^{\epsilon_2-1}{\omega} \\ 0
\end{Bmatrix},
\label{eq:tangentO}
\end{equation}
\begin{equation}
    t_{\eta} = \begin{Bmatrix} -a_1 \epsilon_1  \sin{\eta} \cos{\eta}^{\epsilon_1-1}
   \cos{\omega}^{\epsilon_2} \\
   -a_2 \epsilon_1 \sin{\eta} \cos{\eta}^{\epsilon_1-1}  \sin{\omega}^{\epsilon_2} \\ 
 a_3 \epsilon_1 \cos{\eta} \sin{\eta}^{\epsilon_1-1}
 \end{Bmatrix},
\label{eq:tangentE}
\end{equation}
and define the tangent plane. 
The unit tangent vectors show that the direction of $t_{\omega}$ is independent of $\epsilon_1$, and that $t_{\eta}$ depends on the cross-section parameter $a_{\omega}$ defined in Eq.(\ref{eq:aSubOmega}),
\begin{equation}
        t_{\omega_U} = \begin{Bmatrix} \frac{-a_1 \cos^{\epsilon_2-2}{\omega}}{\sqrt{(a_1\cos^{\epsilon_2-2}{\omega})^2+(a_2\sin^{\epsilon_2-2}{\omega})^2}} \\
        \frac{a_2 \sin^{\epsilon_2-2}{\omega}}{\sqrt{(a_1\cos^{\epsilon_2-2}{\omega})^2+(a_2\sin^{\epsilon_2-2}{\omega})^2}} \\ 0
\end{Bmatrix},
\label{eq:tangentOU}
\end{equation}

\begin{equation}
        t_{\eta_U} = \begin{Bmatrix} \frac{-a_1 \cos^{\epsilon_2}{\omega}}{\sqrt{a_3^2(\cos^{2-\epsilon_1}{\eta}\sin^{\epsilon_1-2}{\eta})^2+a_{\omega}^2}} \\
        \frac{-a_2 \sin^{\epsilon_2}{\omega}}{\sqrt{a_3^2(\cos^{2-\epsilon_1}{\eta}\sin^{\epsilon_1-2}{\eta})^2+a_{\omega}^2}} \\ 
        \frac{a_3}{\sqrt{a_3^2+(\cos^{\epsilon_1-2}{\eta}\sin^{2-\epsilon_1}{\eta})^2 a_{\omega}^2)}}
\end{Bmatrix},
\label{eq:tangentEU}
\end{equation}
We can also see that the tangent is well defined at the limits for the segment considered, but the variation of the tangent will not be continuous when adding the other quadrants or as a function of $\epsilon$; the values can be found in Table \ref{tab:t1} and are shown in Figure \ref{fig:tangentLimitsOmega}. These limit points are important because in some cases they define the maximum and minimum curvature points.

\begin{table}[htbp]
\caption{Tangent vectors at the boundary points (seams)}
\begin{center}
\begin{tabular}{|c|c|c|}
\hline
\multicolumn{3}{|c|}{ $t_{\omega_U}$} \\[2ex]
\hline
\textbf{$\epsilon_2$} & \textbf{$\omega = 0$}& \textbf{$\omega = \frac{\pi}{2}$} \\[1ex]
\hline
 $\epsilon_2 < 2$ & $\begin{Bmatrix} 0 \\ 1 \\ 0 \end{Bmatrix}$ & $\begin{Bmatrix} -1 \\ 0 \\ 0 \end{Bmatrix}$ \\
\hline
 $\epsilon_2 = 2$ & $\begin{Bmatrix} \frac{-a_1}{\sqrt{a_1^2+a_2^2}} \\ \frac{a_2}{\sqrt{a_1^2+a_2^2}} \\ 0 \end{Bmatrix}$ & $\begin{Bmatrix} \frac{-a_1}{\sqrt{a_1^2+a_2^2}} \\ \frac{a_2}{\sqrt{a_1^2+a_2^2}} \\ 0 \end{Bmatrix}$ \\
\hline
$\epsilon_2 > 2$ & $\begin{Bmatrix} -1 \\ 0 \\ 0 \end{Bmatrix}$ & $\begin{Bmatrix} 0 \\ 1 \\ 0 \end{Bmatrix}$ \\
\hline
\multicolumn{3}{|c|}{$t_{\eta_U}$}  \\ [2ex]
\hline
\textbf{$\epsilon_1$} & \textbf{$\eta = 0$}& \textbf{$\eta = \frac{\pi}{2}$} \\[1ex]
\hline
$\epsilon_1 < 2$ & $\begin{Bmatrix} 0 \\ 0 \\ 1 \end{Bmatrix}$ & $\begin{Bmatrix} \frac{-a_1\cos^{\epsilon_2}{\omega}}{a_{\omega}} \\ \frac{-a_2\sin^{\epsilon_2}{\omega}}{a_{\omega}} \\ 0 \end{Bmatrix}$ \\
\hline
$\epsilon_1 = 2$ & $\begin{Bmatrix}  \frac{-a_1\cos^{\epsilon_2}{\omega}}{\sqrt{a_3^2+a_{\omega}^2}} \\ \frac{-a_2\sin^{\epsilon_2}{\omega}}{\sqrt{a_3^2+a_{\omega}^2}} \\ \frac{a_3}{\sqrt{a_3^2+a_{\omega}^2}} \end{Bmatrix}$ & $\begin{Bmatrix} \frac{-a_1\cos^{\epsilon_2}{\omega}}{\sqrt{a_3^2+a_{\omega}^2}} \\ \frac{-a_2\sin^{\epsilon_2}{\omega}}{\sqrt{a_3^2+a_{\omega}^2}} \\ \frac{a_3}{\sqrt{a_3^2+a_{\omega}^2}} \end{Bmatrix}$ \\
\hline
$\epsilon_1 > 2$ & $\begin{Bmatrix}  \frac{-a_1\cos^{\epsilon_2}{\omega}}{a_{\omega}} \\ \frac{-a_2\sin^{\epsilon_2}{\omega}}{a_{\omega}} \\ 0 \end{Bmatrix}$ & $\begin{Bmatrix} 0\\ 0 \\ 1 \end{Bmatrix}$ \\
\hline
\end{tabular}
\label{tab:t1}
\end{center}
\end{table}

\subsection{Normal Vectors and Curvatures}

\begin{figure}[htpb]
      \centering
      \includegraphics[scale=0.18]{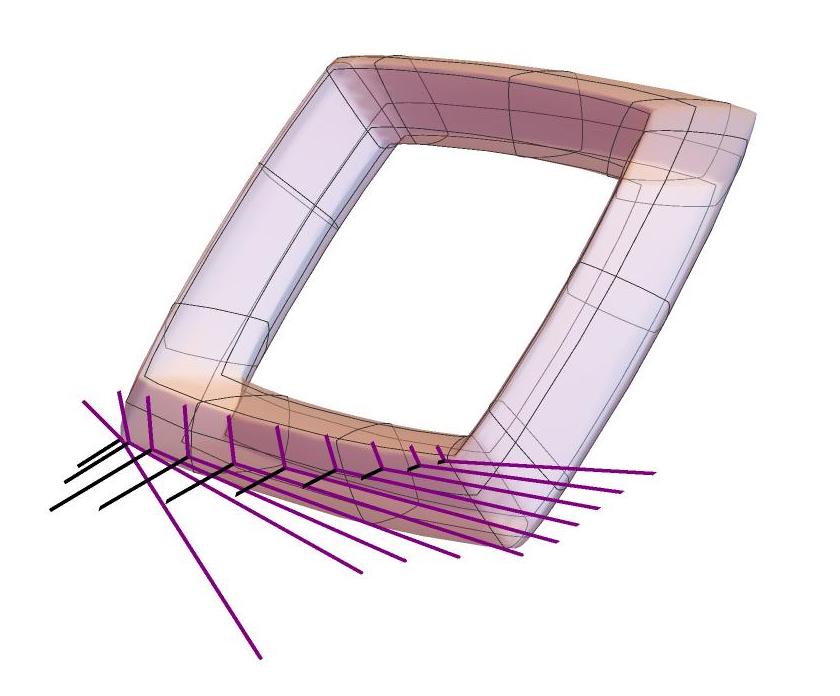}
      \caption{Tangent lines along coordinate curves (purple), and lines normal to the surface (black).}
      \label{fig:normalsST}
   \end{figure}

The tangents along the coordinate curves are used to compute the normal vector to the surface at each point $p$ defined by the Gauss map,
\begin{equation}
   n_p = \frac{1}{m_n}\begin{Bmatrix} a_2 a_3 c_{\eta} (a_4 + 
    c_{\eta}^{\epsilon_1}) c_{\omega} s_{\eta}^{\epsilon_1-1} s_{\omega}^{\epsilon_2-1} \\ 
 a_1 a_3 c_{\eta} (a_4 + 
    c_{\eta}^{\epsilon_1}) c_{\omega}^{\epsilon_2-1}
   s_{\eta}^{\epsilon_1-1} s_{\omega} \\ 
 a_1 a_2 s_{\eta}(a_4 + 
    c_{\eta}^{\epsilon_1})  c_{\omega}^{\epsilon_2-1}c_{\eta}^{\epsilon_1-1} 
    s_{\omega}^{\epsilon_2-1}
   \end{Bmatrix},
\end{equation}
where
\begin{align}
    m_n=&\Big((a_4 + c_{\eta}^{\epsilon_1})^2 ((a_1 a_2 c_{\eta}^ {\epsilon_1-1} s_{\eta} c_{\omega}^{\epsilon_2 - 1} s_{\omega}^{\epsilon_2 - 1})^2 + \nonumber \\
    &(a_3 c_{\eta} s_{\eta}^{\epsilon_1- 1})^2 ((a_1 c_{\omega}^{\epsilon_2 -1} s_{\omega})^2 + (a_2 c_{\omega} s_{\omega}^{\epsilon_2 - 1})^2))\Big)^{1/2}
\end{align}
and $c_{\omega}$ denotes the cosine of $\omega$, similarly for the rest of trigonometric functions. Figure \ref{fig:normalsST} shows a series of tangents along the curves defined by the two parameters, and the corresponding normal to the surface at each point. 


The normal curvatures along the $\omega$ and $\eta$ curves can be calculated using the second fundamental form, 

\begin{align}
   & k_\omega = \frac{ a_1 a_2 \sqrt{(\epsilon_2-2)^2 (a_4 + 
    c_{\eta}^{\epsilon_1})^4 c_{\omega}^{2 \epsilon_2-2}
   s_{\omega}^{2 \epsilon_2-2}}}
   {\epsilon_2( (a_4 + c_{\eta}^{\epsilon_1})^2 c_{\omega}^2 s_{\omega}^2 (a_1^2 c_{\omega}^{2 \epsilon_2-4} + a_2^2 s_{\omega}^{2 \epsilon_2-4}))^{3/2} }, \nonumber \\
   & k_\eta = \frac{a_3 \sqrt{(\epsilon_1-2)^2 c_{\eta}^{2 \epsilon_1-2} s_{\eta}^{2 \epsilon_1-2} (a_1^2 c_{\omega}^{2 \epsilon_2} + a_2^2 s_{\omega}^{2 \epsilon_2})}}
     {\epsilon_1 (c_{\eta}^2 s_{\eta}^2 (a_3^2 s_{\eta}^{2 \epsilon_1-4} + c_{\eta}^{ 2 \epsilon_1-4} (a_1^2 c_{\omega}^{2 \epsilon_2} + 
       a_2^2 s_{\omega}^{2 \epsilon_2})))^{3/2}}.
\label{eq:curvs}
\end{align}
These are not principal curvatures, as the coordinate curves are not perpendicular for most values of coefficients and angles. It can be noted that these values are very near the maximum and minimum values for most supertoroids, and can be used as a first approximation.

At this point, it may be beneficial to select different curvatures depending on the application. One good choice could be to use the  \emph{mean curvature} $H$, defined as the average of the two principal curvatures, $\bar H = \frac{k_1+k_2}{2}$. This can be calculated using the first and second fundamental forms.

The mean curvature may not be the best choice when one direction is preferred, for instance for grasping usign parallel grippers. Similarly, the main curvatures may be calculated using the fundamental forms.  All the expressions defining the differential geometry of the supertoroid are simple enough to be implemented for fast calculations on the surface. 

\section{EXPERIMENTAL RESULTS}

\begin{figure}[h!]
    \centering
    \includegraphics[scale=0.12]{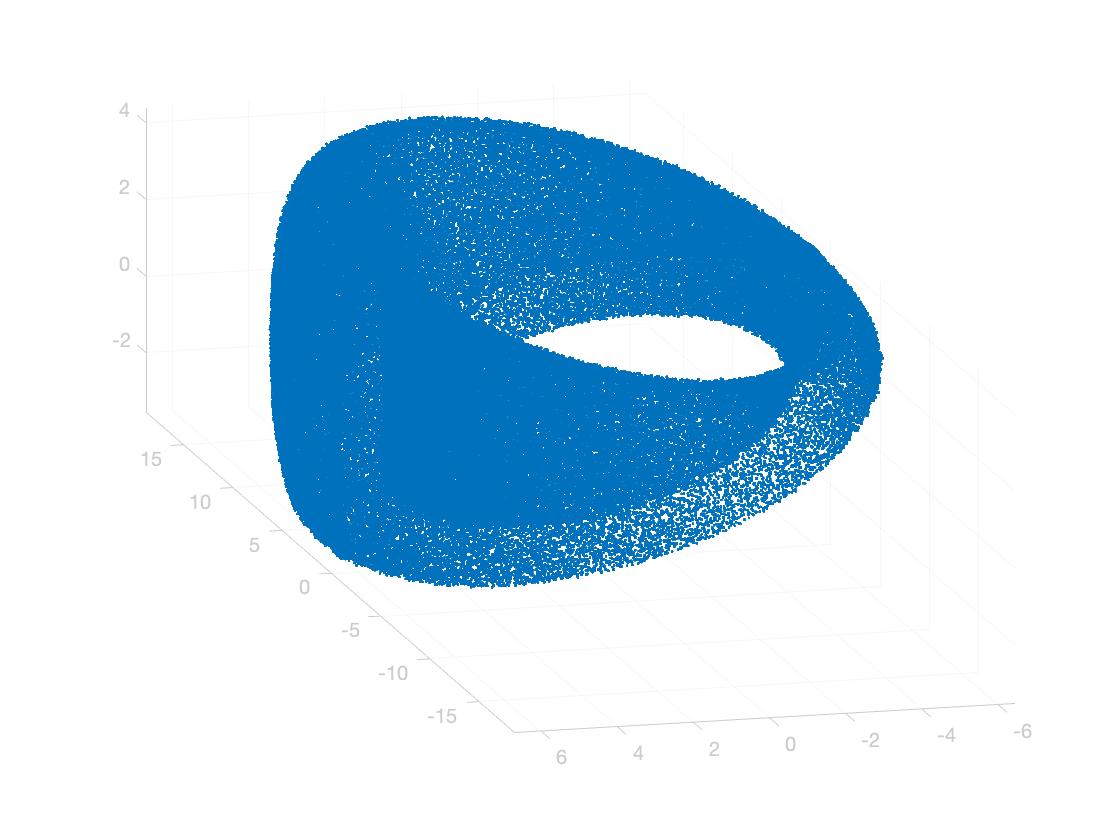}
    \includegraphics[scale=0.50]{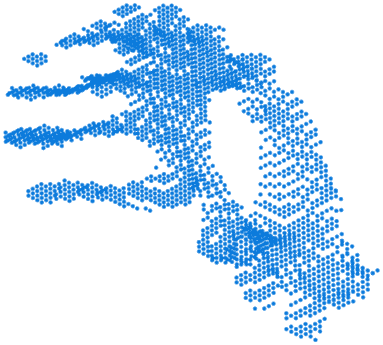}
    \caption{A generated partial, noisy point cloud for a supertoroid (left) and a real partial point cloud showing a helical gear (right). Real cloud obtained using Intel Realsense D435 camera.}
    \label{fig:partialCloud}
\end{figure}

The 3D sensors most commonly used in robotic operations are RGB-D cameras; time-of-flight or structured-light sensors. A single image from one sensor introduces some noise and, more importantly for our application, it obtains only a partial view of the object, from a single viewpoint.  Figure \ref{fig:partialCloud} shows virtual and a real segmented partial clouds. 

\subsection{Partial cloud strategies}

There are three strategies to deal with the partial view issue. The first one is to combine the point clouds of several cameras or several views of a single camera \cite{Nunes:2015}. The second strategy is to complete the point cloud using heuristics. The third strategy is to work with the partial point cloud.

Several strategies have been developed to complete the point cloud, such as \cite{Quispe:2015}, where the shape of the object is completed from the partial view and a mesh of the completed model is generated for grasping.  Generating a full model from a partial view can be approached in several ways, the most common ones being the use of symmetry planes or extrusion  \cite{Beetz:2010} \cite{Peters:2012}. 

While having a complete cloud is the ideal outcome, generating several views increases the time and complexity of operations. The strategies to complete the cloud assume some kind of unknown symmetry, and hence don´t add any additional information to the fitting of superquadrics or supertoroids, which are indeed symmetric. 

In this work, we show that the methodology allows fitting the partial cloud obtained from a single view, skipping the extra step of completing the cloud based on some symmetry assumption and with quick results.

\subsection{Supertoroid fitting}

The fitting of the supertoroid to the point cloud is performed as a least squares problem, minimizing the distance defined in Eq. (\ref{eq:distance}), using an Interior Point method. The parameters to optimize are contained in the vectors $vi = (a_1, a_2, a_3, a_4, \epsilon_1, \epsilon_2)$ defining the shape of the supertoroid, and $v_e = (x, y, z, q_0, q_1, q_2, q_3)$ defining its position -the location of its canonical frame. In total we have 12 independent parameters to define the object.

The experimental procedure shown here is included to highlight additional steps that help in the fitting process. 



\begin{figure}[htpb]
    \centering
    \includegraphics[scale=0.11]{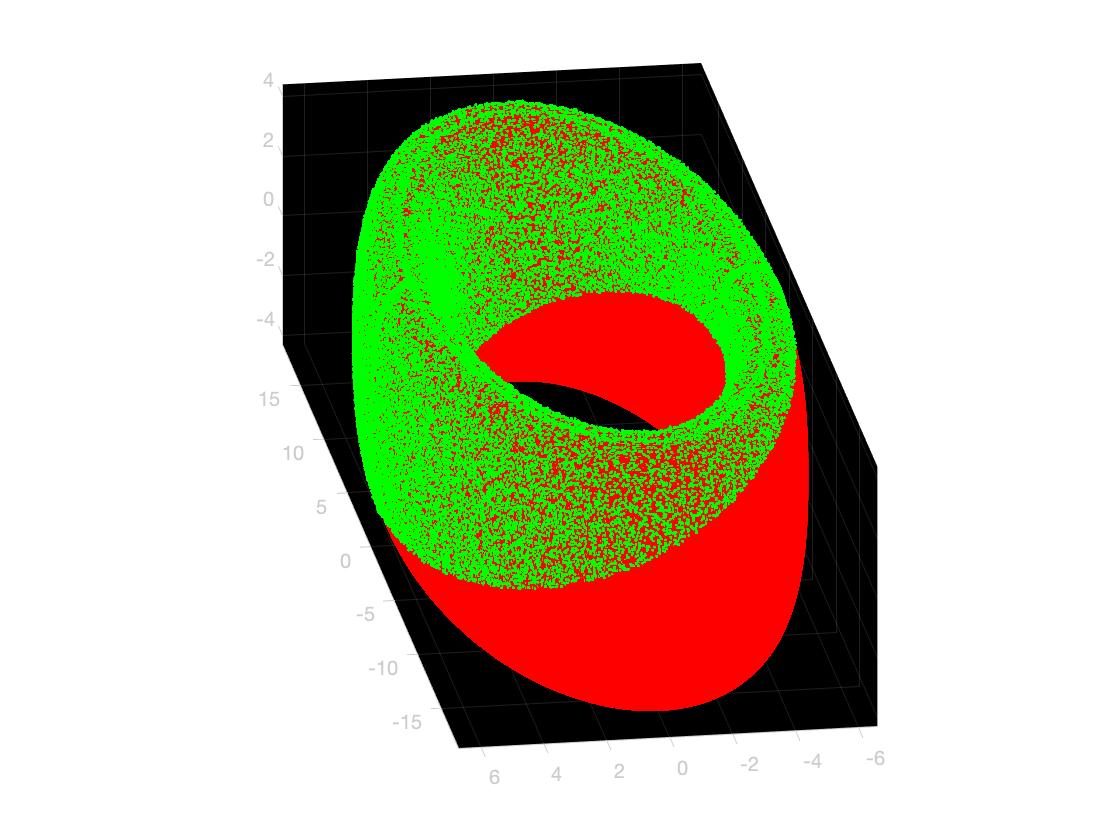}
    \includegraphics[scale=0.10]{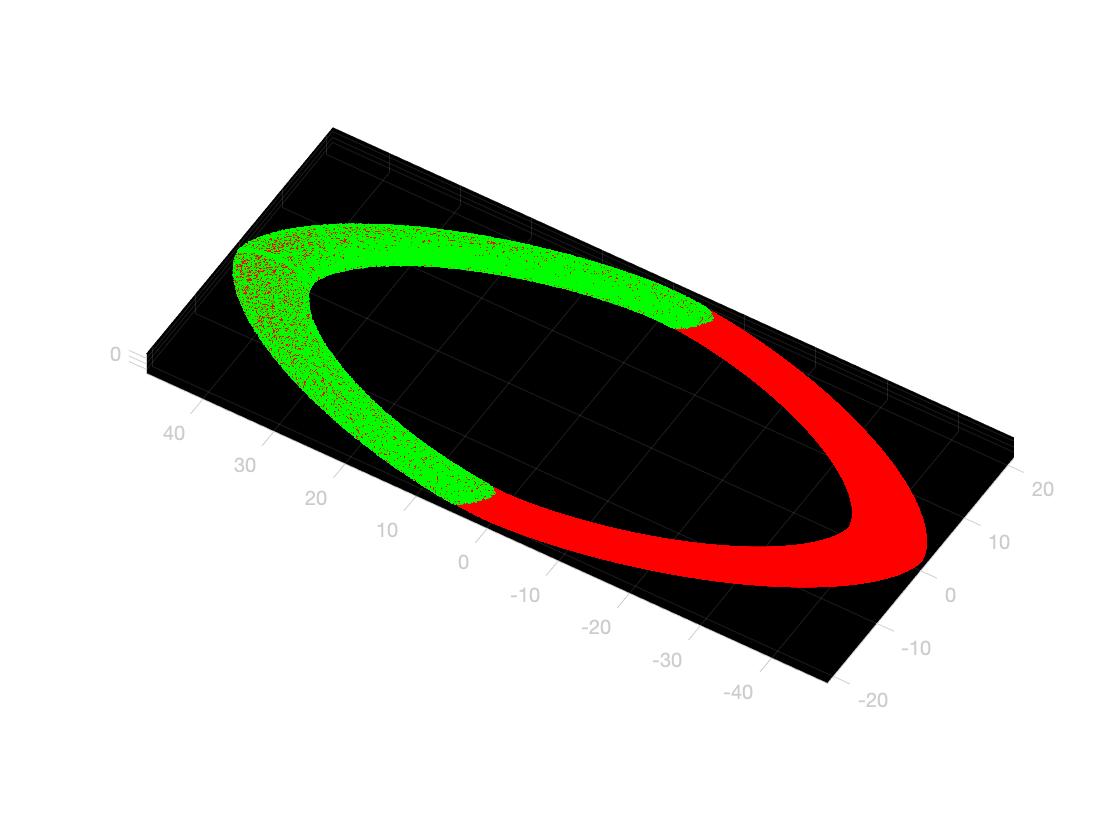}
    \includegraphics[scale=0.10]{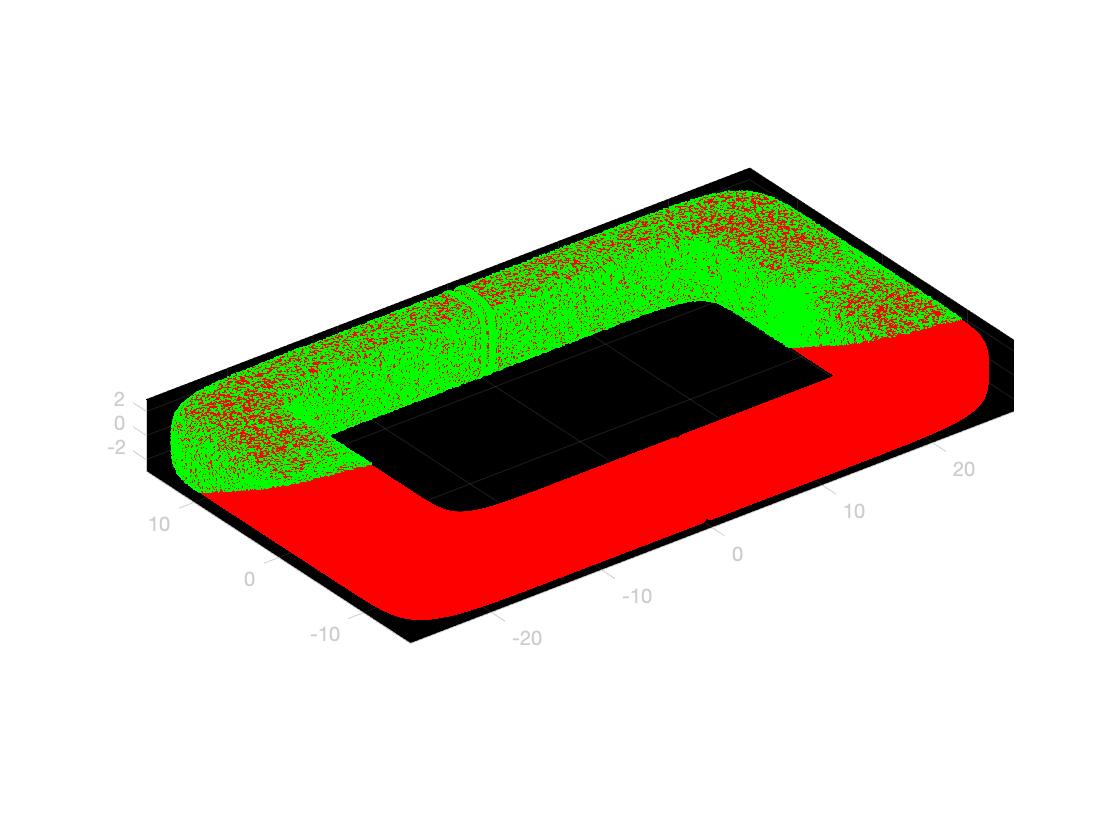}
    \caption{Three examples of supertoroid fitting of centered, partial clouds with random 0.1 noise. Green corresponds to generated cloud, and red corresponds to fitted supertoroid.}
    \label{fig:resultsCentered}
\end{figure}

Before the optimization stage, an initial guess must be provided which will yield good convergence. Principal Component Analysis (PCA) is usually attempted with superquadrics. However, for supertoroids we need to align the hole with the $z$-axis, and the direct application of PCA does not provide this alignment. With the correct alignment of the axes, the fitting yields fast and accurate results, as seen in Figure \ref{fig:resultsCentered}. 

For generally-positioned objects, we find the location and approximate direction of the hole of the supertoroid using the mean superellipse. The variables to be optimized in this stage are the intrinsic parameters $\epsilon_1$, $a_1$, $a_2$, and $a_4$, and the extrinsic parameters.



\begin{figure}[h!]
    \centering
    \includegraphics[scale=0.45]{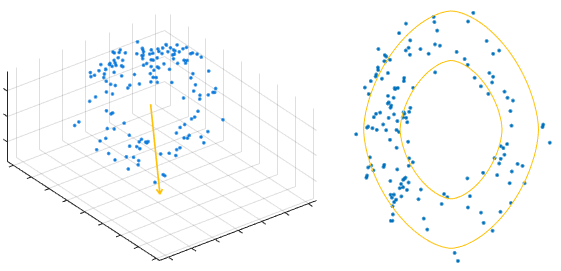}
    \caption{Result of the first optimization stage, showing the downsampled point cloud.}
    \label{fig:stage1}
\end{figure}

An example of the result of this process can be seen in Figure \ref{fig:stage1}. The figure shows the vector normal to the plane of the mean superellipse. Also shown is the projection of the point cloud onto said plane, with the superellipses corresponding to a top-down view of the supertoroid superimposed. With this, we obtain a good approximation of the $z$ axis, corresponding to the hole.

This first stage is run three times, starting from vectors parallel to the $x$, $y$, and $z$ axes. The lowest-cost solution is then used as the initial guess for the second stage.

\begin{figure}[h!]
    \centering
    \includegraphics[scale=0.25]{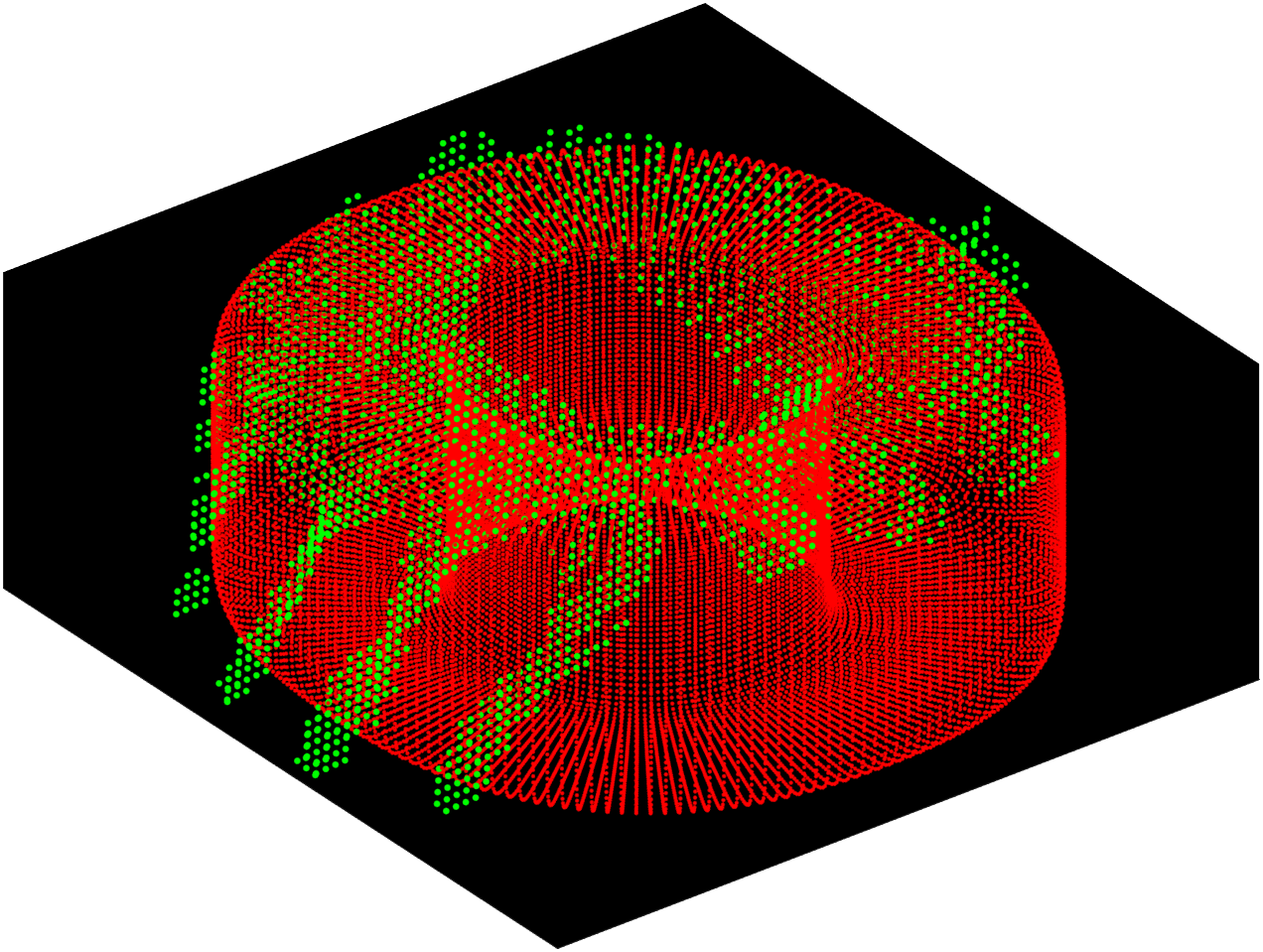}
    \caption{Result with the fitted supertoroid being shown in red and the original point cloud in green.}
    \label{fig:stage2}
\end{figure}

In the second optimization stage, the meridian radial distance is minimized. For partial clouds, it is useful to add a term to maximize $a_4$. The final result for the gear point cloud can be seen in Figure \ref{fig:stage2}.

Note that for both stages the point cloud is downsampled using a random downsampling method. This serves two purposes. The first is to reduce the time per iteration; the second, to make the cost function smoother, thus avoiding some local optima. For the objects studied, the best results were produced using 150 points in the first stage and 1000 in the second.

\subsection{Implementation and combined results}

\begin{figure}[h!]
    \centering
    \includegraphics[scale=0.2]{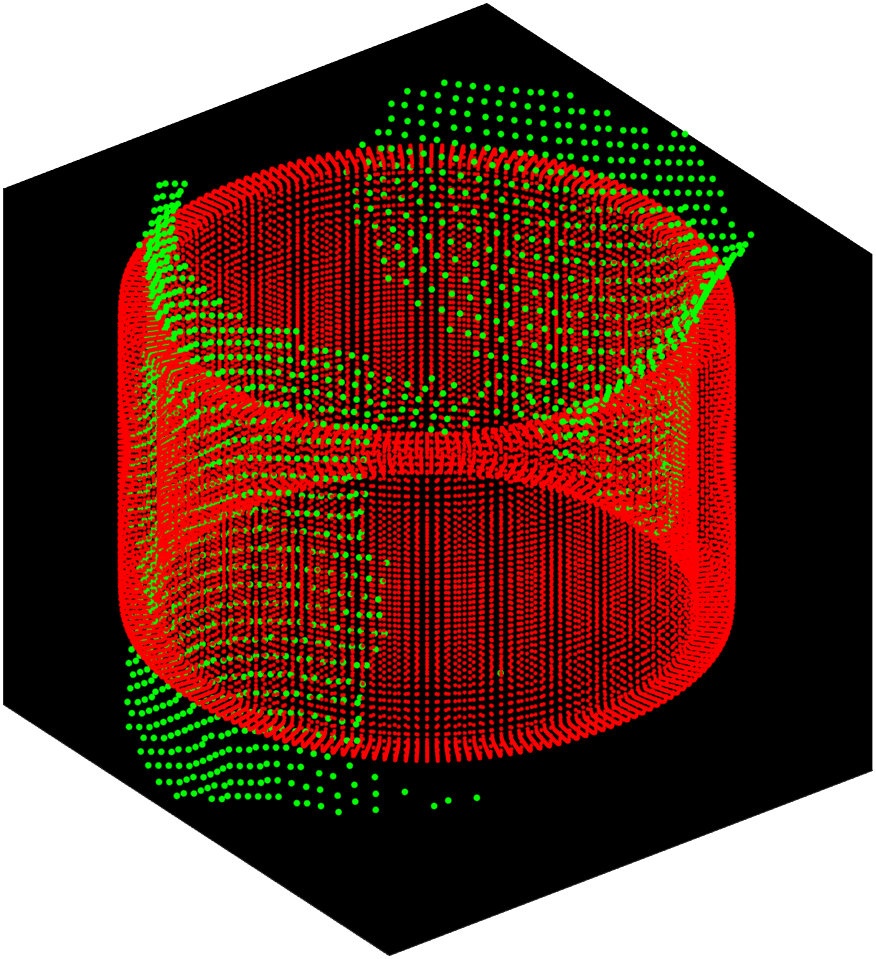}
    \includegraphics[scale=0.2]{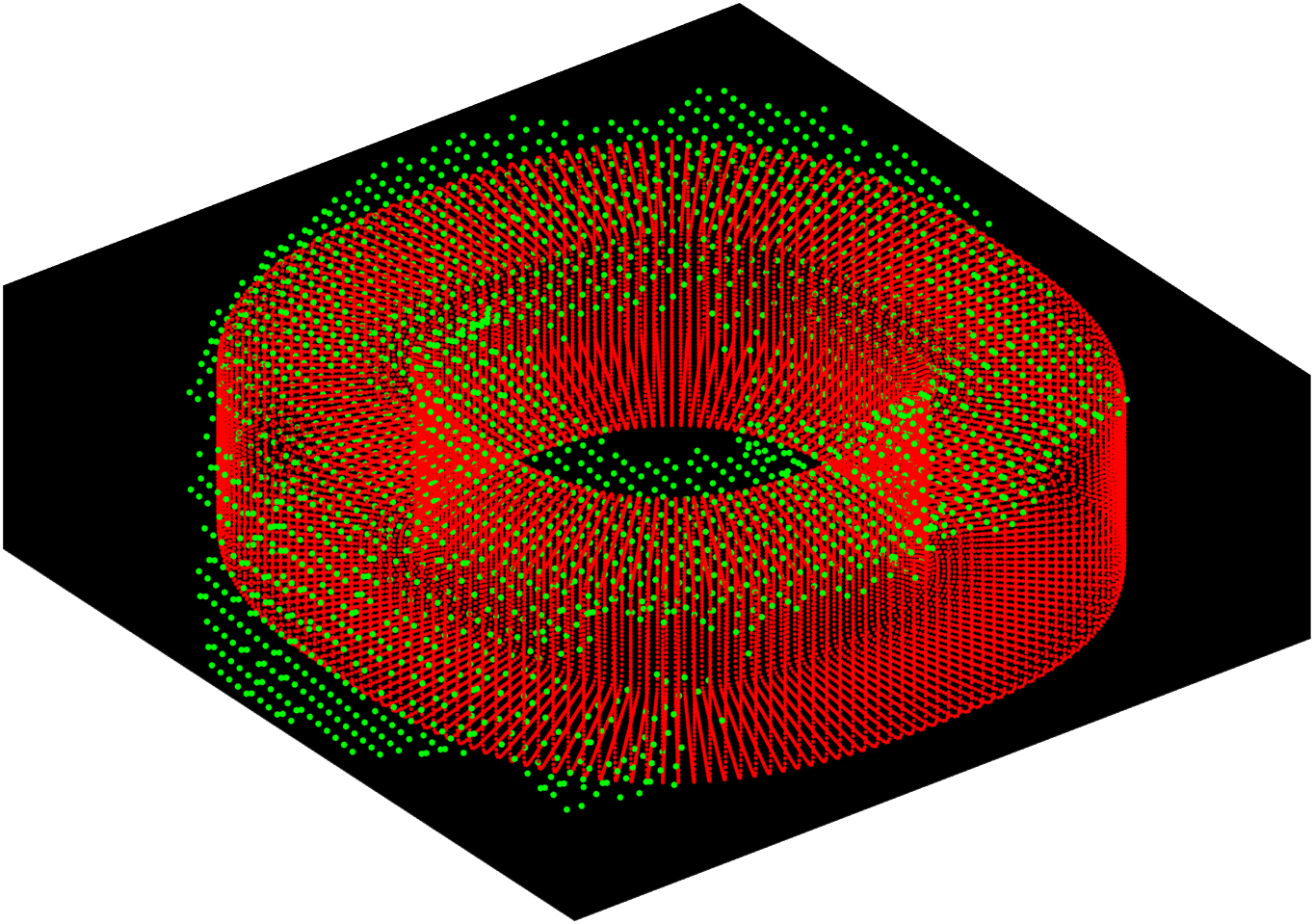}
    \includegraphics[scale=0.2]{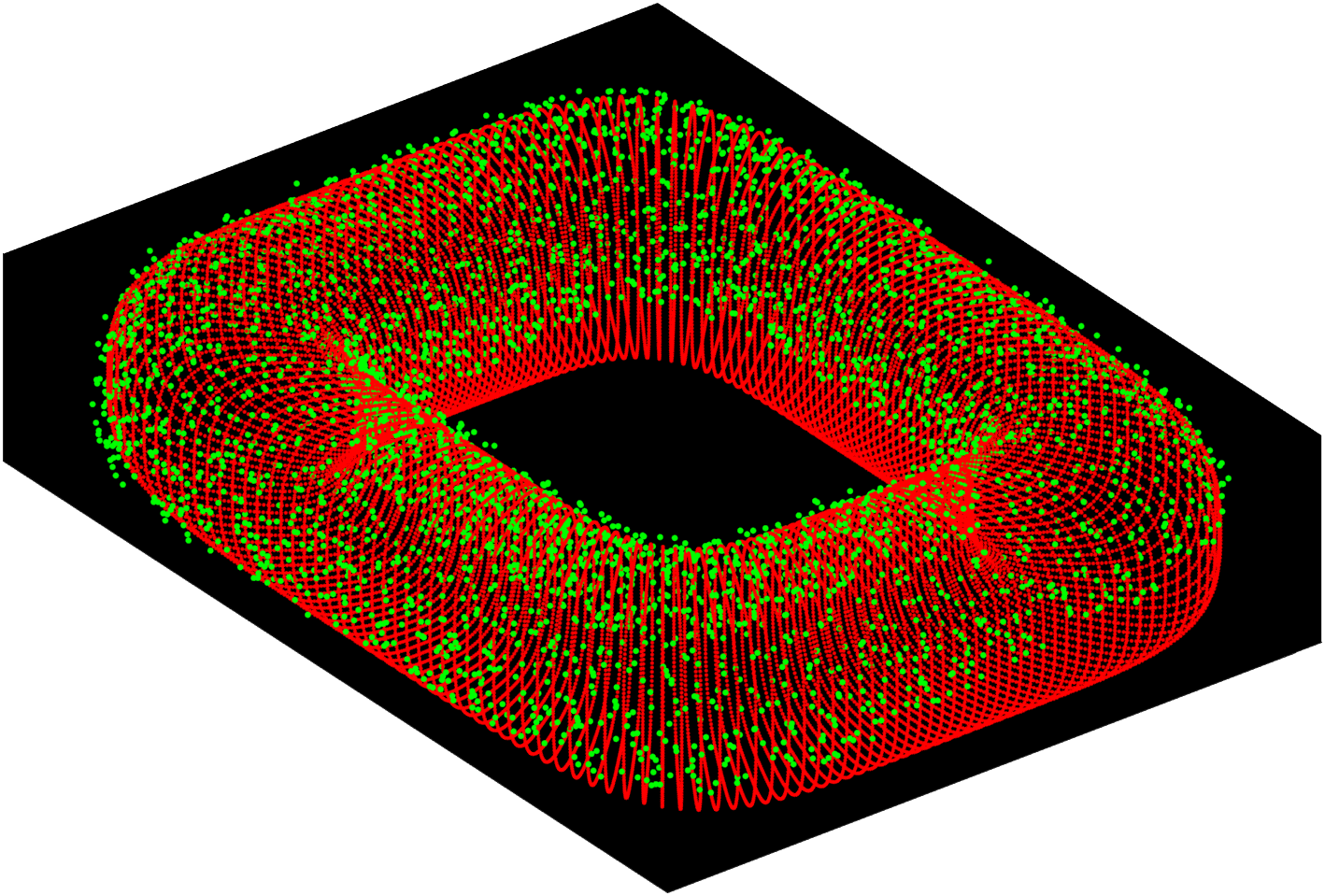}
    \caption{Result of three additional point cloud fittings, the supertoroid is shown in red and the original point clouds in green. The point clouds are "pot" (top left), "bearing" (top right), and "synthetic" (bottom).}
    \label{fig:cloudresults}
\end{figure}

To test and refine the fitting algorithm, it was used on several real and synthetic point clouds. For each point cloud, the algorithm was run 10 times, each time with a different random downsampling. Figure \ref{fig:cloudresults} shows the supertoroids fitted to point clouds of relatively short objects.

\begin{figure}[h!]
    \centering
    \includegraphics[scale=0.2]{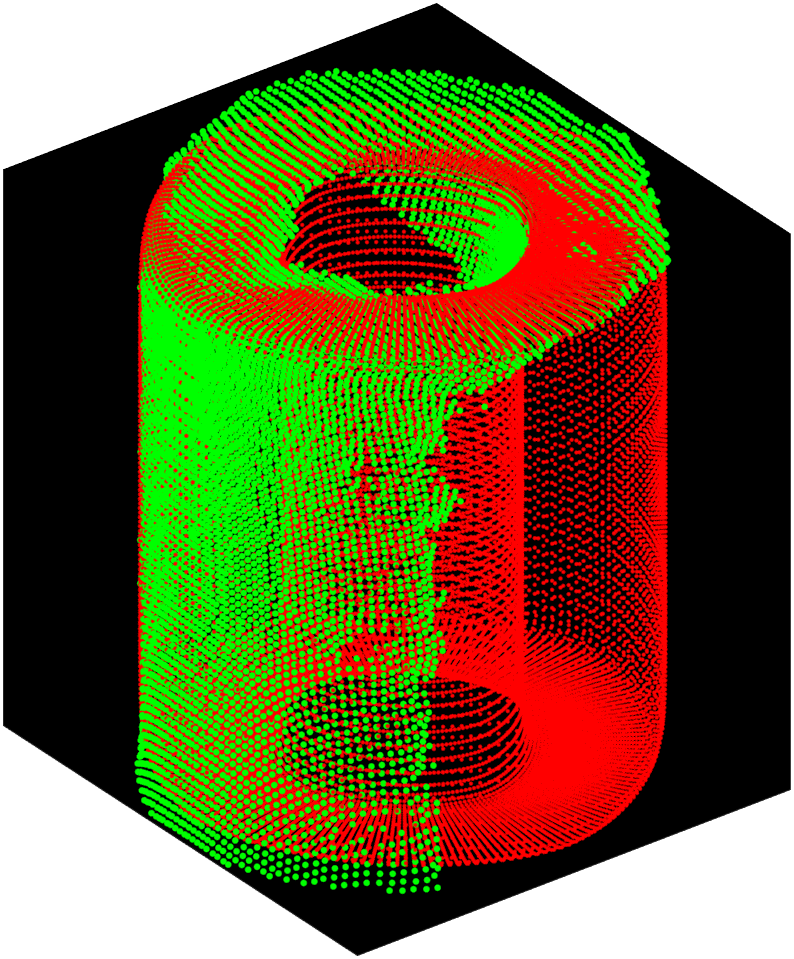}
    \includegraphics[scale=0.2]{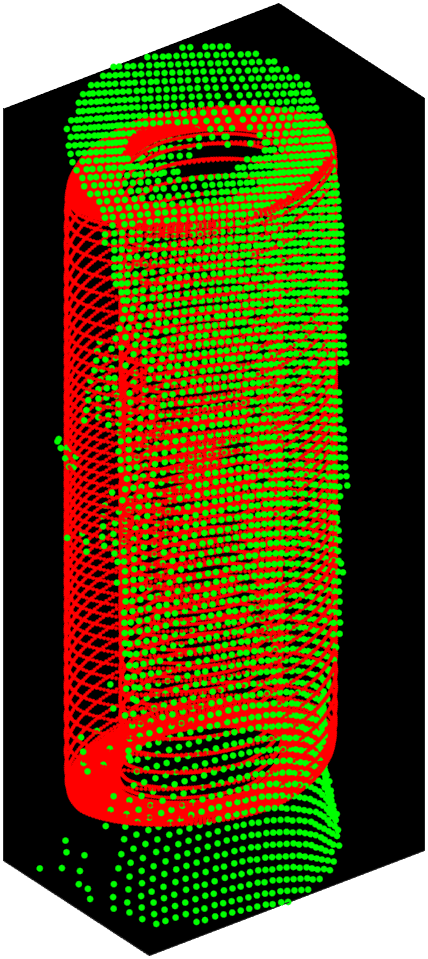}

    \caption{Results of the point clouds of taller objects. The point clouds are "roll" (left) and "long roll" (right).}
    \label{fig:cloudresultstall}
\end{figure}

To get reliable results for narrower objects, the initial guess for $a_3$ needs to be adjusted. This is the case for the solutions shown in Figure \ref{fig:cloudresultstall}. 

The results of the execution time for the fitting process, in an Intel Core i7-10750H CPU@2.60GHz using MATLAB R2023b, are presented in Table \ref{tab:results}. In Table \ref{tab:results}, "G" refers to good results, that is, fits such as the ones shown in Figures \ref{fig:stage2}-\ref{fig:cloudresultstall}. "D" refers to decent fits, which are good enough for grasping in most cases but that present some issues, such as sharp corners where they shouldn't be, holes substantially smaller than that of the real object, etc. "B" refers to bad fits, unsuitable for grasping. The rest of the columns give information on the CPU time required to solve the optimization problems. $t_{S1}$, $t_{S2}$, and $t_{total}$ denote the average time it took to complete stage 1, stage 2, and both, respectively. $s_{total}$ refers to the standard deviation in total time. The average CPU time of all stage 1 problems shown in table \ref{tab:results} is 1.72 seconds with a standard deviation of 0.21 seconds. For the second stage, it is 0.99 seconds with a standard deviation of 0.40 seconds. For the sum of the two, the average is 2.71 seconds with a standard deviation of 0.44 seconds.

\begin{table}[H]
\caption{Fitting results}
\label{tab:results}
\begin{center}
\begin{tabular}{|l|c|c|c|c|c|c|c|}
\hline
Point cloud & G & D & B & $t_{S1}$ [s]& $t_{S2}$ [s]& $t_{total}$ [s]& $s_{total}$ [s]\\
\hline
\hline
Gear & 8 & 1 & 1 & 1.65 & 1.04 & 2.70 & 0.54\\
\hline
Pot & 8 & 0 & 2 & 1.83 & 1.11 & 2.95 & 0.58\\
\hline
Bearing & 10 & 0 & 0 & 1.82 & 1.06 & 2.87 & 0.22\\
\hline
Synthetic & 10 & 0 & 0 & 1.78 & 0.57 & 2.35 & 0.23 \\
\hline
Roll & 8 & 1 & 1 & 1.59 & 0.95 & 2.54 & 0.36\\
\hline
Long roll & 6 & 4 & 0 & 1.64 & 1.19 & 2.83 & 0.39\\
\hline
\end{tabular}
\end{center}
\end{table}



Stage 1 is only necessary when the shape and location of the object are unknown. Therefore, most of the time only stage 2 will be executed, with the first stage only being called when track is lost of, or first viewing, an object.

Additionally, information on the pose (e.g. on a known flat surface, hole facing up) and proportions (especially $a_3$) of the object should be incorporated into the initial guesses when available to increase the reliability of the results. 

The algorithms have been implemented in Matlab code, available at \url{https://github.com/jbadiat/SupertoroidFitting}. The point clouds of objects with holes, of different shapes and sizes, have been captured using a RealSense 435 camera.

\section{CONCLUSIONS}

In this work we study the supertoroid and introduce the meridian radial distance, derived following the radial distance used for superquadrics. We also present the differential geometry of the supertoroid surface. Both can be used for the grasping of unknown objects with holes, including partial, noisy point clouds obtained from the computer vision of holed objects. Supertoroids can take a variety of shapes, from square to circular to rhomboid, in order to adapt to a variety of objects. The identification of the supertoroid position and geometry, based on 12 parameters, yields a very simple kinematic definition for shape and pose identification and allows creating grasping heuristics for the object.  
The results show that shape and pose identification of unknown objects with the newly-developed meridian radial distance is computationally fast and somewhat robust to noise and partial knowledge of the object.



\section*{ACKNOWLEDGEMENT}
This work has been partially supported by Agencia Estatal de Investigación under project PID2020-117509GB-I00/AEI/10.13039/50110001103.

\bibliographystyle{IEEEtran.bst}

\bibliography{arxiv}

\end{document}